\documentclass[11pt]{article}
\usepackage[a4paper,left=1.7cm,right=1.7cm,top=2cm,bottom=2cm]{geometry}
\usepackage{setspace}  
\usepackage{amsmath,amssymb,bm}
\usepackage{graphicx}
\usepackage{siunitx}
\usepackage[numbers,super,sort&compress]{natbib}
\usepackage{xcolor}
\usepackage[hidelinks]{hyperref}
\usepackage{caption}
\usepackage{float}
\usepackage{booktabs} 

\usepackage{comment}
\usepackage{subcaption} 
\usepackage{tabularx}   
\usepackage{multirow}   
\usepackage{bbm}        
\usepackage{siunitx}    

\begin{document}

\begin{titlepage}
\centering
\vspace*{1cm}
{\LARGE\bfseries
A Bias-Corrected Two-Stage Approach for Joint Modelling of Multidimensional Longitudinal HRQoL \\and Survival Data\par}

\vspace{1.2cm}

{\large
Hortense Doms\par
LIDAM/ISBA, UCLouvain, Louvain-la-Neuve, Belgium\par
\vspace{0.3cm}
and\par
\vspace{0.3cm}
Philippe Lambert\par
Institut de Math\'ematiques, Universit\'e de Li\`ege, Belgium\par
LIDAM/ISBA, UCLouvain, Louvain-la-Neuve, Belgium\par
\vspace{0.3cm}
and\par
\vspace{0.3cm}
Catherine Legrand\par
LIDAM/ISBA, UCLouvain, Louvain-la-Neuve, Belgium\par
}

\vspace{1.4cm}

{\bfseries Abstract\par}
\vspace{0.4cm}

\begin{minipage}{0.85\textwidth}
\setstretch{1.0} 
Health-related quality-of-life (HRQoL) outcomes are increasingly incorporated into oncology research to complement traditional survival endpoints by capturing patients’ well-being over time. These outcomes are typically collected through multidimensional questionnaires yielding longitudinal ordinal data, and are often subject to dropout due to disease progression or death. In this context, joint models provide a well-established framework to account for the dependence between longitudinal HRQoL trajectories and time-to-event outcomes, but fully joint estimation rapidly becomes computationally prohibitive when multiple latent dimensions and random effects are involved. We propose a novel slope-corrected two-stage (SC2S) approach for the joint analysis of multivariate ordinal HRQoL data and survival outcomes within a multidimensional latent trait framework. The proposed approach propagates longitudinal information to the survival model through informative priors on the random effects, while additionally re-estimating longitudinal slope parameters. This strategy substantially reduces bias in both longitudinal and survival submodels while preserving much of the computational efficiency of two-stage procedures. Through simulation studies and an application to HRQoL data from patients with progressive glioblastoma, we show that the proposed method closely approximates fully joint Bayesian estimation while requiring notably less computation time.
\end{minipage}

\vspace{0.6cm}

\begin{minipage}{0.85\textwidth}
\end{minipage}

\end{titlepage}

\setstretch{1.3}

\section{Introduction}\label{sec1}
In oncology clinical research, overall survival (OS) is traditionally regarded as the primary endpoint for assessing the benefit of new treatments. Although OS remains a key measure of therapeutic efficacy, it does not fully reflect how treatments affect patients’ daily functioning and well-being. Consequently, complementary patient-centered endpoints have emerged, particularly health-related quality of life (HRQoL). HRQoL is typically assessed using patient-reported questionnaires administered repeatedly throughout treatment and follow-up, offering valuable insight into patients’ perceptions of their physical, psychological, and social health \citep{fayers2013}. Patient-reported outcomes such as HRQoL thus provide essential complementary information, contributing to a more comprehensive and patient-centered assessment of treatment benefit. Their importance has also motivated the development of advanced statistical methods to appropriately analyze these complex longitudinal data.\\

\noindent HRQoL questionnaires are typically composed of multiple scales, each assessed through several items. Responses are often recorded on Likert-type scales (e.g., “not at all,” “a little,” “quite a bit,” “very much”), producing ordinal data. A widely used method for analyzing such data is the scoring approach \citep{Fayers2001}, in which item responses are summed to obtain a total score for each scale and time point. However, this method presents important limitations \citep{GorterFox2015}: it ignores heterogeneity in response patterns that yield the same total score, leading to loss of information, and the resulting scores are frequently treated as continuous despite their ordinal nature and potential ceiling or floor effects.\\

\noindent To overcome these issues, item response theory (IRT) methods, and in particular graded response (GRM) models, have been increasingly adopted for HRQoL analysis \citep{barbieri}. These models estimate a latent variable that represents the unobserved true health status using observed item responses regarded as manifestations of this latent construct. To capture the correlation structure inherent to longitudinal data, multilevel IRT extensions have been proposed, allowing the latent trajectory to depend on covariates and subject-specific random effects. More recently, the unidimensional assumption has been relaxed by introducing multidimensional latent trait mixed models \citep{Wang2017}, enabling multiple latent dimensions and, when relevant, within-item multidimensionality.\\

\noindent These longitudinal data are frequently analysed in the presence of intercurrent or terminal events. In clinical trials, patients may experience events such as death or loss to follow-up, which interrupt the collection of HRQoL measurements. The occurrence of such events is often related to the underlying health status, resulting in dependence between the event process and the longitudinal outcomes. Ignoring this informative dropout may therefore lead to biased estimates \citep{henderson}. In this context, the joint analysis of the event hazard and longitudinal data, known as the joint modelling framework, has become increasingly common. \citet{Touraine} recently highlighted the importance of joint models for HRQoL data when informative dropout is present. \citet{Saulnier} extended the joint modelling methodology to handle repeated data from measurement scales using a cumulative probit model and cause-specific hazard functions. \citet{Doms} proposed an approach to handle informative dropout in longitudinal HRQoL data through a joint modelling framework, which integrates a graded response model for multivariate ordinal outcomes and a cause-specific proportional hazards model for informative and non-informative dropout events. Both of these approaches are restricted to a unidimensional latent trait. However, \citet{Wang2019} introduced a joint multidimensional latent trait mixed model with a proportional hazards submodel in a Bayesian framework. Because they rely on fully joint estimation, all these methods are computationally intensive, particularly when several latent dimensions and random effects are involved. Indeed, joint models for multivariate longitudinal data often face substantial computational burden, as the number of random effects grows with the number of outcomes, leading to slow estimation algorithms \citep{papageorgiou2019overview}.\\

\noindent Early approaches to the estimation of joint models relied on two-stage methods \citep{self, tsiatis1995}. In these approaches, the longitudinal submodel is fitted first, and its estimates are subsequently used in the survival model. Although computationally appealing, such procedures may produce biased estimates under informative dropout \citep{tsiatis2004joint, ye2008}. For this reason, full joint estimation was later proposed. This approach corrects bias by estimating both submodels simultaneously, but it is substantially more computationally demanding. As joint models have evolved to accommodate increasingly complex longitudinal and survival structures, their computational burden has grown accordingly. To address these challenges, some authors have proposed bias-corrected two-stage methods, which retain the computational advantages of two-stage estimation while reducing bias. \citet{Mauff2020} introduced importance-sampling corrections to reduce bias in multivariate joint models, while \citet{alvares2023} extended the Bayesian two-stage framework with informative priors for random effects and bias-adjustment techniques, improving accuracy while maintaining computational efficiency. Both of these approaches are developed within the context of exponential-family distributions for the longitudinal responses.\\

\noindent In this work, we adapt the corrected two-stage approach proposed by \citet{alvares2023} to efficiently estimate a joint model combining a multidimensional latent trait mixed model for multiple longitudinal ordinal outcomes and a proportional hazards model with a flexible baseline hazard, within a Bayesian framework. In addition, unlike \citet{alvares2023}, we propose a correction for the longitudinal slope parameters, mitigating the bias that may arise when the longitudinal and survival components are fitted separately. Our goal is to approximate the accuracy of fully joint estimation while significantly reducing the computational cost. \\

\noindent The rest of this paper is organized as follows. Section~\ref{sec2} introduces the proposed joint model. Section~\ref{sec3} presents the Bayesian estimation strategy based on the corrected two-stage approach. Section~\ref{sec4} reports a simulation study assessing the method’s performance, and Section~\ref{sec5} applies our approach to HRQoL data from the QLQ-BN20 questionnaire completed by glioblastoma patients. Section~\ref{sec6} concludes the paper.

\section{Methodology}\label{sec2}
\subsection{Longitudinal submodel}\label{long}
\noindent
We consider a sample of size $n$ patients followed over time. For each individual $i = 1, \dots, n$, we observe $K$ longitudinal ordinal items, where item $k$ $(k = 1, \dots, K)$ has $L_k$ ordered categories. The longitudinal measurements are denoted by $y_{ik}(t)$, corresponding to item $k$ observed at time $t = t_{ij}$ $(j = 1, \dots, N_i)$. All items are coded such that larger values indicate worse clinical conditions.  

\noindent
We assume the existence of $P$ latent variables, also called latent dimensions, with $P < K$, representing distinct abilities for each subject. The vector of latent processes for individual $i$ at time $t$ is denoted as $\bm{\eta}_{i}(t) = \left(\eta_{i}^{(1)}(t), \dots, \eta_{i}^{(p)}(t), \dots, \eta_{i}^{(P)}(t)\right)^\top$.
We consider that each item loads on one or more latent dimensions and that higher values of $\eta_{i}^{(p)}(t)$ correspond to a lower ability level for dimension $p$.

\subsubsection{Latent process}
\noindent Latent variables cannot be directly observed but can be assessed through the longitudinal responses provided to the different items. The trajectory of $\bm{\eta}$ over time for patient $i$ is described using a linear mixed model. Specifically, for the $p$th latent variable $(p = 1,\dots,P)$, we have :  
\normalsize

    \begin{equation*}
	\left\lbrace
	\begin{array}{l}
	\eta^{(p)}_{i}(t) =  \bm{X_i}^{(p)}(t) \, \bm{\beta}^{(p)} + \bm{Z_i}^{(p)}(t) \, \bm{b_i}^{(p)} \\
	\bm{b_i} \sim \mathcal{N}(0,\Sigma) 
	\end{array}\right.
	\end{equation*} 

 \noindent 
where $\bm{X_i}^{(p)}(t)$ and $\bm{Z_i}^{(p)}(t)$ are the covariates associated to the fixed and random effects for each latent variable ${\eta}^{(p)}_{i}(t)$. The vector $\bm{b_i} = (\bm{b_i}^{(1)}, \dots, \bm{b_i}^{(p)}, \dots, \bm{b_i}^{(P)})^\top$ contains all the normally distributed individual random effects. The correlations between latent variables are captured through the covariance structure of $\bm{b_i}$. The fixed effects term describes the average trajectory of the latent process at the population level, while the random effects term captures individual-specific deviations from this trajectory. 

\subsubsection{Graded response model}

\noindent 
We now define the link between the latent process $\bm{\eta}_i(t)$ and each observed longitudinal ordinal response using the graded response model (GRM) \citep{samejima1969estimation}. The cumulative probability for the $k$th ordinal outcome, with $L_k$ ordered categories,
is defined for $l = 1,\dots,L_k-1$ as:
\normalsize
\begin{equation*}
\begin{aligned}
\mathrm{logit}\left\{ \mathrm{P}(y_{ik}(t) \leq l \mid \bm{\eta}_i(t)) \right\}
&= d_{kl} - \bm{a}_k^\top \bm{\eta}_i(t)\\[0.3cm]
\quad \Leftrightarrow \quad
\mathrm{P}(y_{ik}(t) \leq l \mid \bm{\eta}_i(t)) 
&= \frac{1}{1 + \exp\{-(d_{kl} - \bm{a}_k^\top \bm{\eta}_i(t))\}}.
\end{aligned}
\end{equation*}

\noindent
Each item $k$ is characterized by a vector of discrimination parameters 
$\bm{a}_k = (a_k^{(1)}, \dots, a_k^{(P)})^\top$ of dimension $P$, and by a set of threshold parameters 
$\bm{d}_k = (d_{k1}, \dots, d_{k,L_k-1})^\top$. 
The discrimination vector $\bm{a}_k$ quantifies the strength of association between item $k$ and each latent dimension. 
A larger absolute value of $a_k^{(p)}$ indicates that item $k$ is more sensitive to differences along latent dimension $p$. The threshold parameter $d_{kl}$ determines, together with $\bm{a}_k$, the location in the latent space where the probability of responding in category $l$ or below equals 50\%. 
Specifically, this occurs for latent variable vectors $\bm{\eta}_i(t)$ satisfying $\bm{a}_k^\top \bm{\eta}_i(t) = d_{kl}.$
The thresholds must satisfy the ordering constraint 
$d_{k1} < d_{k2} < \dots < d_{k,L_k-1}$. The probability of observing a response in category $l = 1,\dots,L_k$ is obtained from the cumulative probabilities as
\[
\mathrm{P}(y_{ik}(t) = l \mid {\eta}_i(t)) 
= \mathrm{P}(y_{ik}(t) \le l \mid {\eta}_i(t)) 
- \mathrm{P}(y_{ik}(t) \le l-1 \mid {\eta}_i(t)),
\]
with the standard conventions
\begin{equation*}
\mathrm{P}(y_{ik}(t) \le 0 \mid {\eta}_i(t)) = 0,
\qquad
\mathrm{P}(y_{ik}(t) \le L_k \mid {\eta}_i(t)) = 1.
\end{equation*}

\noindent In multidimensional graded response models, identifiability constraints are required because the latent traits are not directly observed. Without such constraints, the model is invariant to translations, rescalings, and rotations of the latent space, meaning that different combinations of thresholds, discrimination parameters, and latent variables can produce exactly the same response probabilities. These constraints ensure identifiability by fixing, for each dimension, a reference origin and unit, and by preventing the different dimensions from overlapping. To achieve this, several identification strategies are possible, but a common approach consists in selecting, for each latent dimension, an anchor item whose parameters are constrained to fix the scale, location, and orientation of the latent space. This typically involves fixing one discrimination parameter per dimension (e.g., $a_k^{(p)} = 1$), setting one threshold to a reference value (e.g., $d_{k1} = 0$), and restricting cross-loadings for these anchor items.

\subsection{Survival process}\label{surv}

\noindent Let $T_i^*$ denote the time to a terminal event for the $i$-th individual (e.g., death or disease progression). We observe $T_i = \min(T_i^*, C_i)$, where $C_i$ is the administrative (and non-informative) right censoring time (end of study). We model the time-to-event using a proportional hazard model 
\begin{equation*}
 h_{i}(t)  = h_{0}(t) \exp \big\{\bm{\gamma}^\top\bm{w}_i  + \bm{\alpha}^\top m( t, \bm{b_i}) \big \} ,  \,  \, \, \, t >0 
\end{equation*}

\noindent where $h_{0}(t)$ denotes the baseline hazard and $\bm{w}_i$ is the vector of baseline survival covariates for subject $i$ with regression coefficients $\bm{\gamma}$. Parameters $\alpha$ quantify the association between the longitudinal latent process and the survival process. Different association functions $m(t, \bm{b_i})$ can be specified. 
For example, the current-value $m(t, \bm{b_i}) = \bm{\eta}_i(t)$ and the random-effects $m(t, \bm{b_i}) = \bm{b_i}$ structures are the most frequently used in the literature \citep{rizopoulos2012joint}, although other specifications are also possible. \\

\noindent In the joint modelling framework, the baseline hazard should not be left unspecified; therefore we define the logarithm of the baseline hazard functions using B-splines \citep{Lambert2005, rizopoulos2016r} :  
$$\log \{ h_0(t) \} = \sum_{u=1}^{U} \gamma_{h_0,u} B_u(t),
\qquad
\bm{\gamma}_{h_0} = (\gamma_{h_0,1}, \dots, \gamma_{h_0,U})^\top,
$$
\noindent where $\gamma_{h_{0},u}$ is the regression coefficient associated to the $u$th function of a large B-spline basis associated to equidistant knots on $(0, \max_i T_i)$.

\section{Bayesian inference}\label{sec3}
We apply a Bayesian approach to estimate our proposed joint model that links a time-to-event outcome and a multidimensional latent trait linear mixed model, defined by
\begin{equation}\label{eq1}
\left\lbrace
\begin{array}{l}
 h_{i}(t)  = h_{0}(t) \exp \big\{\bm{\gamma}^\top\bm{w}_i + \bm{\alpha}^{\top} m( t, \bm{b_i}) \big \},\\[0.2cm]
\eta^{(p)}_{i}(t) =  \bm{X_i}^{(p)}(t) \, \bm{\beta}^{(p)} + \bm{Z_i}^{(p)}(t) \, \bm{b_i}^{(p)}  \\[0.2cm]
\mathrm{logit}\lbrace \mathrm{P}(y_{ik}(t) \leq l | \bm{\eta}_{i}(t)) \rbrace = d_{kl} -  \bm{a_{k}^\top} \bm{\eta}_{i}(t)
\end{array}\right.
\end{equation}

\subsection{Posterior distribution}
As introduced in Section~\ref{sec1}, the model in \eqref{eq1} may be estimated using a full joint modelling approach or via two-stage strategies. Each approach leads to a different formulation of the posterior distribution, which we detail in the following subsections. For each approach, the posterior distribution is explored using Markov Chain Monte Carlo (MCMC) methods.

\subsubsection{Joint specification (JS)}
\noindent In the framework of joint models for longitudinal and survival data, the expression for the posterior distribution of the model parameters is derived under the assumption that the longitudinal and survival processes are independent conditionally on the random effects $\bm{b_i}$. Moreover, the longitudinal measurements $\bm{y_i}$ of each individual are assumed independent given the random effects. Therefore, the joint posterior distribution is defined by 
\begin{equation}\label{eq2}
p(\bm{\theta}, \bm{b} \mid \bm{T}, \bm{\delta}, \bm{y}) \propto
\prod_{i=1}^n
p(T_i, \bm{\delta}_i \mid \bm{b}_i; \bm{\theta}_s)
\prod_{j=1}^{N_i} \prod_{k=1}^{K}
 p\!\left(y_{ik}(t_{ij}) \mid \bm{b}_i; \bm{\theta}_y \right)
p(\bm{b}_i; \bm{\theta}_b)\, p(\bm{\theta}).
\end{equation}
where $\bm{\theta}$ denotes the full parameter vector with joint prior $p(\bm{\theta})$. Specifically, $\bm{\theta} = [\bm{\theta_s}^\top, \bm{\theta_y}^\top, \bm{\theta_b}^\top]^\top$ with $\bm{\theta_s}$ for the survival submodel, $\bm{\theta_y}$ for the longitudinal submodel, and $\bm{\theta_b}$ for the random-effects covariance structure. The expressions for $p(T_i, \bm{\delta}_{i} \, | \,\bm{b_i}; \bm{\theta})$, $ p\!\left(y_{ik}(t_{ij}) \mid \bm{b}_i; \bm{\theta}_y \right)$ and $p(\bm{b_i} ; \bm{\theta_b}) $ can be found in Web Appendix S1.

\subsubsection{Standard Two-Stage (S2S) and Corrected Two-Stage (C2S) approaches}\label{C2S}
Exploration of the joint posterior distribution in Equation~\eqref{eq2} using MCMC is computationally intensive and may suffer from convergence issues due to the complex hierarchical structure of joint models. Some authors have proposed avoiding these limitations by using two-stage estimation strategies, as introduced in Section~\ref{sec1}.\\

\noindent In the \textit{Standard Two-Stage} (S2S) approach, the longitudinal and survival submodels are estimated sequentially. In the first stage, the longitudinal submodel is fitted independently, providing posterior estimates (often posterior means) for the parameters $\bm{\theta_y}$ and individual random effects $\bm{b_i}$, denoted $\hat{\bm{\theta}}_y$ and $\hat{\bm{b}}_i$. In the second stage, these estimated quantities are inserted as fixed covariates into the survival submodel, through $m_i(t \mid \hat{\bm{b}}_i, \hat{\bm{\theta}}_y)$, to infer on $\bm{\theta_s}$. While computationally attractive, this approach ignores the informative dropout mechanism in the first step, leading to biased estimates. Using these biased predictors in the survival model further propagates this bias, especially for the association parameter. Moreover, by treating random effects ($\hat{\bm{b}}_i$) as fixed values, it implicitly assumes they are known without error in the second stage. Such an omission fails to propagate the sampling variability from the first stage, causing a significant loss in the variability of the random effects and leading to an artificial reduction in the uncertainty of the survival estimators \citep{tsiatis2004joint, ye2008}.\\

\noindent
To address these issues, \citet{alvares2023} proposed a \textit{Corrected Two-Stage} (C2S) approach, which provides a practical compromise between the S2S and JS approaches. As in the first stage of the S2S approach, the longitudinal submodel is fitted separately to obtain $\hat{\bm{\theta}}_y$ and $\hat{\bm{\theta}}_b$. In the second stage, instead of fixing $\bm{b_i}$ to their point estimates after the first stage, the posterior distribution of $(\bm{\theta_s}, \bm{b})$ is derived using the likelihood of the joint model, with the longitudinal parameters and the variance--covariance matrix of the random effects treated as known and replaced by their estimates from the first stage:
\begin{equation}\label{eq3}
p(\bm{\theta_s}, \bm{b} \mid \bm{T}, \bm{\delta}, \bm{y}, \hat{\bm{\theta}}_y, \hat{\bm{\theta}}_b) 
\propto 
\prod_{i=1}^{n} p(T_i, {\delta}_i \mid \bm{b}_i; \bm{\theta_s}) 
\, \prod_{j=1}^{N_i} \prod_{k=1}^{K}
 p\!\left(y_{ik}(t_{ij}) \mid \bm{b}_i; \hat{\bm{\theta}}_y \right)
\, p(\bm{b}_i; \hat{\bm{\theta}}_b) 
\, p(\bm{\theta_s}).
\end{equation}

\noindent
In this formulation, the random effects $\bm{b}_i$ are treated as individual-level parameters with informative prior distributions $p(\bm{b}_i; \hat{\bm{\theta}}_b)$ obtained from the first stage, where $\hat{\bm{\theta}}_b = \hat{\bm{\Sigma}}$. This reduces the model hierarchy by removing the integration over random effects, thereby improving convergence speed and computational efficiency compared to a full joint specification. Moreover, incorporating the longitudinal likelihood function in the second stage acts as a bias correction mechanism, allowing part of the longitudinal uncertainty to be reflected in the survival component. By specifying informative prior distributions based on the first-stage results, the C2S approach allows the random effects to be re-estimated as random variables within the survival submodel rather than fixed covariates. This strategy effectively accounts for the inherent variability of the latent trajectories during the second stage, ensuring that the estimated survival parameters exhibit a posterior dispersion comparable to the fully joint model.\\

\noindent
Overall, it turns out that the corrected two-stage approach maintains most of the accuracy of full joint modeling while substantially reducing the computational burden and improving stability in complex settings.

\subsubsection{Slope-Corrected Two-Stage (SC2S) Approach}

The C2S approach proposed by \citet{alvares2023} reduces bias in the survival parameters and random effects compared with the standard two-stage (S2S) method. However, in its first step, the C2S approach still ignores the dependence between the longitudinal measurements and the event process. The longitudinal submodel is estimated without accounting for subject dropout due to the occurrence of the event, although this information may be informative and should be taken into account.  \\ 

\noindent The occurrence of the event may strongly affect the observed longitudinal trajectories: subjects at higher risk tend to experience the event earlier and therefore contribute fewer measurements. By fitting the longitudinal model separately in Step~1, two-stage approaches fail to account for this selection mechanism and may lead to biased estimation of the longitudinal time trend \citep{tsiatis2004joint}. \\

\noindent This phenomenon was highlighted by \citet{Mauff2020}, who showed that two-stage and fully joint (JS) estimators differ substantially at the random-effects level, particularly for the random slopes, whereas the random intercepts show much smaller differences. This suggests that time-related random effects are particularly prone to bias under the two-stage approach. Since the population-level slope parameters (denoted $({\beta}_2^{(p)})_{p=1}^P$ in \eqref{eq4}) reflect the average evolution across individual trajectories, their estimation may be affected by bias in the random slopes. \\

 \noindent As discussed in Section~\ref{C2S}, the C2S approach partially corrects the bias in the random effects during Step~2 by re-estimating them jointly with the survival parameters. However, the other longitudinal parameters are kept fixed at their Step~1 estimates. In particular, the potential bias affecting the slope parameters $({\beta}_2^{(p)})_{p=1}^P$ is not corrected. \\

 \noindent To address this limitation, we propose an extension of the C2S method, referred to as the \textit{Slope-Corrected Two-Stage} (SC2S) approach, which allows the longitudinal fixed (population-level) slopes $(\beta_2^{(p)})_{p=1}^P$ to be jointly updated in the second stage, together with the random effects and survival parameters. By re-estimating the slope parameters under the joint likelihood instead of relying exclusively on the longitudinal submodel, the SC2S approach incorporates information from both the longitudinal and survival processes and allows correction of the bias introduced at the first stage, when these parameters were estimated separately.\\
 
\noindent The posterior distribution in \eqref{eq3} is therefore modified as follows:
\begin{equation*}
\begin{aligned}
p(\bm{\theta}_s, \bm{\theta}_{y,t}, \bm{b} \mid \bm{T}, \bm{\delta}, \bm{y}, 
   \hat{\bm{\theta}}_{y,0}, \hat{\bm{\theta}}_b)
\;\propto\;
&\prod_{i=1}^{n}
p(T_i, \delta_i \mid \bm{b}_i; \theta_s) \prod_{j=1}^{N_i} \prod_{k=1}^{K}
p\!\left(y_{ik}(t_{ij}) \mid \bm{b}_i, \bm{\theta}_{y,t}; \hat{\bm{\theta}}_{y,0} \right) \\[0.25cm]
\times\;&p(\bm{b}_i; \hat{\bm{\theta}}_b)\,
p(\bm{\theta}_s)\,
p(\bm{\theta}_{y,t}) \\[0.05cm]
\end{aligned}
\end{equation*}
where $\bm{\theta}_y = (\bm{\theta}_{y,0}, \bm{\theta}_{y,t})$, 
with $\bm{\theta}_{y,t} = (\beta_2^{(p)})_{p=1}^P$ denoting the longitudinal population-level slope parameters updated in the second stage, and $\bm{\theta}_{y,0}$ collecting the remaining longitudinal parameters,
which are kept fixed at their first-stage estimates $\hat{\bm{\theta}}_{y,0}$.

\subsection{Priors distributions}\label{priors}
\noindent
We specify weakly informative prior distributions for the full vector of model parameters $\theta$ \citep{papageorgiou2019overview,Alsefri}. These priors are used consistently across all estimation strategies (JS, SC2S, C2S and S2S) to ensure comparability of inference. \\ 

\noindent  Independent normal priors are assigned to the regression coefficients $\bm{\beta}$, $\bm{\gamma}$, and $\bm{\alpha}$. All free discrimination parameters $\bm{a}$ are assigned independent normal priors. Several strategies can be used to impose order constraints on the thresholds. Here, we reparameterize them through their successive first-order differences and place weakly informative half-normal priors  $\mathcal{N}^+(0,10)$ on these differences, which ensures that the thresholds remain ordered. \\

\noindent Rather than assigning a prior directly to the random-effects covariance matrix \(\Sigma\), we adopt a decomposition into marginal variances and a correlation matrix. Two alternative specifications can be considered. First, the variances of the random effects are assigned inverse-gamma \(\text{IG}(0.01,0.01)\) priors, while the corresponding correlation parameters follow uniform distributions on \([-1,1]\). An alternative specification consists of using an LKJ (Lewandowski\allowbreak–\allowbreak Kurowicka\allowbreak–\allowbreak Joe) prior \citep{lkj2009} for the correlation matrix \({R}\), which shrinks correlations toward zero unless strongly supported by the data, together with weakly informative priors (e.g., half-normal) on the standard deviations $(\bm{\sigma}_b)$. The covariance matrix is then parameterized as $\Sigma = \mathrm{diag}(\bm{\sigma}_b)\,{R}\,\mathrm{diag}(\bm{\sigma}_b)$. These prior choices are standard in the literature and have been shown to provide stable and robust estimation of covariance structures. \\

\noindent The flexibility of the log-baseline hazard, expressed as a linear combination of multiple B-spline basis functions, is regularized using a Gaussian Markov field prior on $\bm{\gamma_{h_0}}$ \citep{eilers1996flexible,Lambert2005}:
\begin{equation*}
p(\bm{\gamma_{h_0}} \mid \tau)  \propto \tau^{\rho(\bm{K})/2} 
\exp\!\left(-\frac{\tau}{2} \bm{\gamma_{h_0}}^\top \bm{K} \bm{\gamma_{h_0}} \right),
\quad \text{with} \quad 
\tau \sim \text{Gamma}(\mathfrak{a},\mathfrak{b}),
\end{equation*}
where $\tau$ is a roughness penalty parameter and $\bm{K} = \bm{\Delta_r}^\top \bm{\Delta_r}$ is the penalty matrix of rank $\rho(\bm{K}) = U - r$ associated with the baseline hazard function $h_0$, where $\bm{\Delta}_r$ denotes the $r$th difference penalty matrix (typically $r=2$). The amount of smoothness is controlled by $\tau$, with $\mathfrak{a}=1$ and $\mathfrak{b}$ set to a small value (e.g., 0.005) \citep{lang2004bayesian}.
\section{Simulation studies}\label{sec4}
We conducted a simulation study to assess the performance of the corrected two-stage approaches, considering both estimation accuracy and computational cost, within the framework of the multivariate latent trait joint model defined in \eqref{eq1}. This simulation study had two primary aims. First, we evaluated to what extent the corrected two-stage approaches (C2S and SC2S) improve estimation accuracy relative to the standard two-stage approach (S2S), while preserving the computational efficiency of S2S. Second, we evaluated the estimation quality of the SC2S approach by analysing the bias, RMSE, and coverage probabilities for all parameters of the joint model.

\subsection{Simulation design}
We generated $500$ datasets, each including $n = 500$ units with up to seven scheduled measurement times. We considered $K = 10$ ordinal outcomes, each with $L_k = 5$ categories. The maximum follow-up time was set to $6$. Longitudinal ordinal data were generated from the latent process (see Section~\ref{long})
\begin{equation}
    \label{eq4}
    \eta^{(p)}_{i}(t) = \beta_0^{(p)} + \beta_1^{(p)} x_i + \beta_2^{(p)} t + b_{i0}^{(p)} + b_{i1}^{(p)} t, \qquad p \in \{1,2\},
\end{equation}
and the event process (see Section~\ref{surv}) followed
\[
h_i(t) = h_0(t) \exp \Big\{ \gamma x_i + \sum_{p=1}^2 \big( \alpha_0^{(p)} b_{i0}^{(p)} + \alpha_1^{(p)} b_{i1}^{(p)} \big) \Big\}.
\]
We refer to $\alpha_0^{(1)}$ and $\alpha_0^{(2)}$ as the random-intercept association parameters, and to $\alpha_1^{(1)}$ and $\alpha_1^{(2)}$ as the random-slope association parameters. The binary covariate $x_i$ was simulated from $\mathrm{Bernoulli}(0.5)$ to mimic a balanced treatment assignment. Regression coefficients were set to
\[
\bm{\beta}^{(1)} = (0.5, -1, 0.75)^\top, \qquad
\bm{\beta}^{(2)} = (1, -0.5, 0.75)^\top, \qquad
\bm{\gamma} = 0.5.
\]

\noindent Random effects $\bm{b}_i = (b_{i0}^{(1)}, b_{i1}^{(1)}, b_{i0}^{(2)}, b_{i1}^{(2)})^\top$
were drawn from a multivariate normal distribution
$\mathcal{N}(\bm{0}, \Sigma)$, where the covariance matrix $\Sigma$ is defined element-wise by
\[
\Sigma_{rs} = \rho_{rs}\,\sigma_r\,\sigma_s, \qquad r,s = 1,\dots,4,
\]
with $\sigma_r^2 = 1$ for all $r$, and $\rho_{rs} = \rho_{sr}$ denoting the correlation between the
$r$th and $s$th components of $\bm{b}_i$.
Specifically, we set
$\rho_{12} = 0.30$, $\rho_{13} = 0.20$, $\rho_{14} = 0.10$,
$\rho_{23} = 0.10$, $\rho_{24} = 0.20$, and $\rho_{34} = 0.30$.\\

\noindent For identifiability, we set $d_{11} = d_{21} = 0$, $a_1^{(1)} = a_2^{(2)} = 1$, and $a_1^{(2)} = 0$. Values of the discrimination and threshold parameters are reported in Table S1 in Supplementary Section S2. Across the 500 simulated datasets, the censoring rate was 10.8\%, with a median event time of 2.28 (interquartile interval: 1.09--4.11). Each subject contributed on average 3.27 longitudinal measurements.\\

\noindent In addition, we considered an alternative simulation scenario based on a current-value association structure,
\[
m(t, \bm{b}_i) = \big( \eta^{(1)}(t), \eta^{(2)}(t) \big).
\]
Parameter values for the event and latent processes under this scenario are reported in Supplementary Section S3. The discrimination and threshold parameters were kept identical to those used in the random-effects association structure.\\

\noindent MCMC samples for the different models were generated using Stan via the R interface \texttt{cmdstanr} (version~2.36.0), relying on the No-U-Turn Sampler (NUTS), an adaptive Hamiltonian Monte Carlo (HMC) algorithm, in Stage~1, and using MCMC algorithms implemented in \textsf{R} for Stage~2. To accelerate computation during Stage~1, the evaluation of the longitudinal likelihood was parallelised over three CPU cores using Stan's \texttt{reduce\_sum} procedure. Note that all two-stage approaches share the same first-stage estimation procedure.

\subsection{Simulation results}
Table \ref{tab1} summarises the computation time for each method. All two-stage methods substantially reduced runtime relative to the fully joint model (JS), with S2S being the fastest as expected. 
\begin{table}[!h]
 \sf\centering
\caption{Computation time (in minutes) per dataset}
\label{tab1}
\begin{tabular}{lccc}
\hline
  & Mean & Median & Central 95\% interval \\
  \hline
JS   & 21.6 & 21.1 & 15.4 -- 31.8 \\
SC2S & 12.3 & 12.0 &  \;\,8.8 -- 18.5 \\
C2S  & 11.8 & 11.5 &  \;\,8.3 -- 18.0   \\
S2S  & 11.3 & 11.0 &  \;\,7.8 -- 17.5 \\
\hline
\end{tabular}
\end{table}

\begin{figure}[H]
    \centering
        \includegraphics[width=1\linewidth]{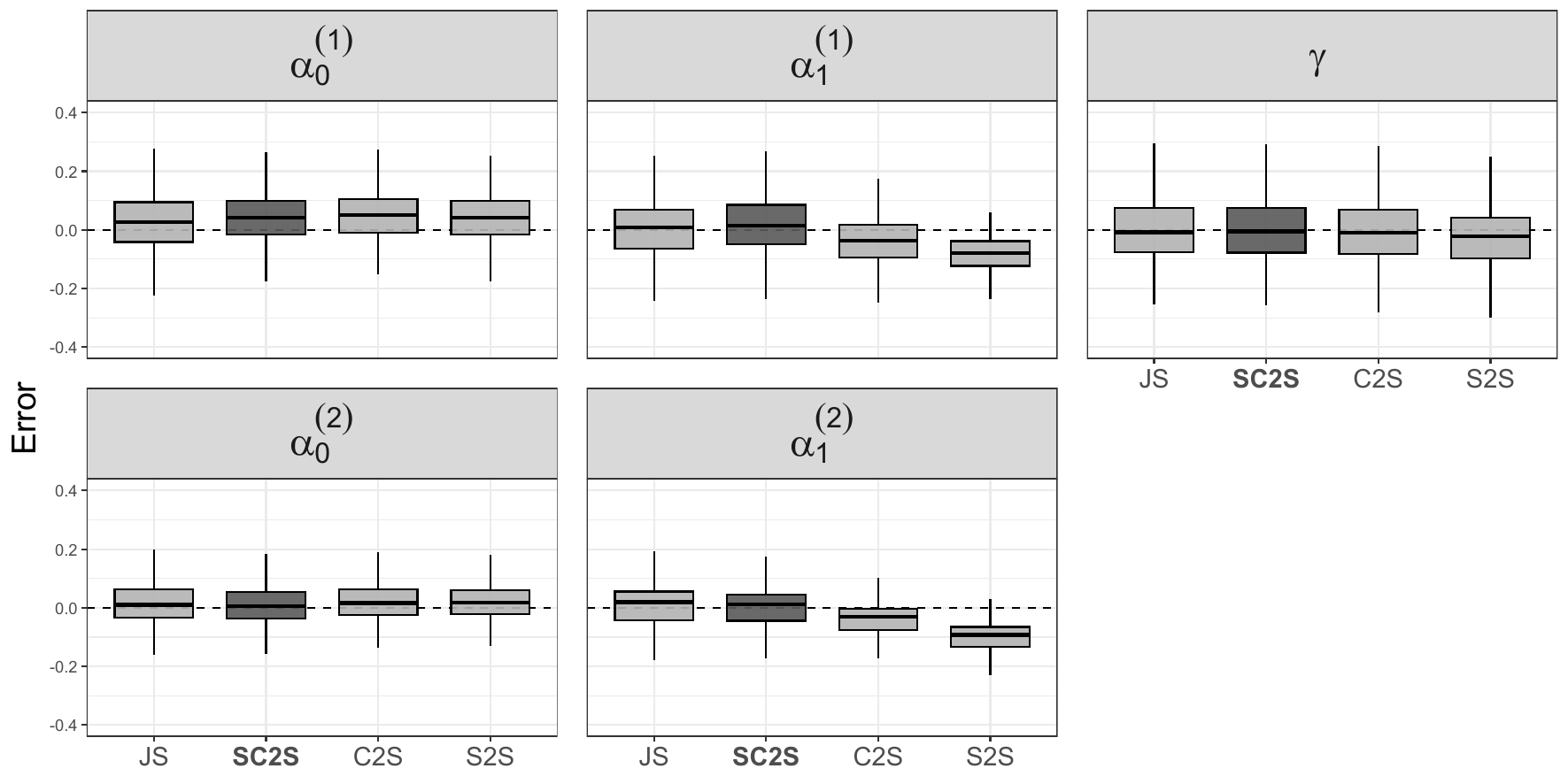}
    \caption{Estimation errors for the survival parameters}
    \label{fig1}
\end{figure}

\noindent Figure~\ref{fig1} presents the estimates of the survival coefficients and the association parameters. As anticipated, S2S shows larger bias, particularly for the random-slope association parameters ($\alpha_1^{(1)}$ and $\alpha_1^{(2)}$), whereas bias for the random-intercept association parameters is relatively small. This is consistent with the fact that random intercepts are less influenced by the dropout mechanism. By re-estimating the random effects in the second stage, C2S partially reduces this bias. Our proposed SC2S method provides the largest improvement: for slope-association parameters, SC2S biases were $0.02$ and $-0.01$, compared with $-0.09$ and $-0.11$ under S2S. \\

\begin{figure}[H]
    \centering
     \includegraphics[width=1\linewidth]{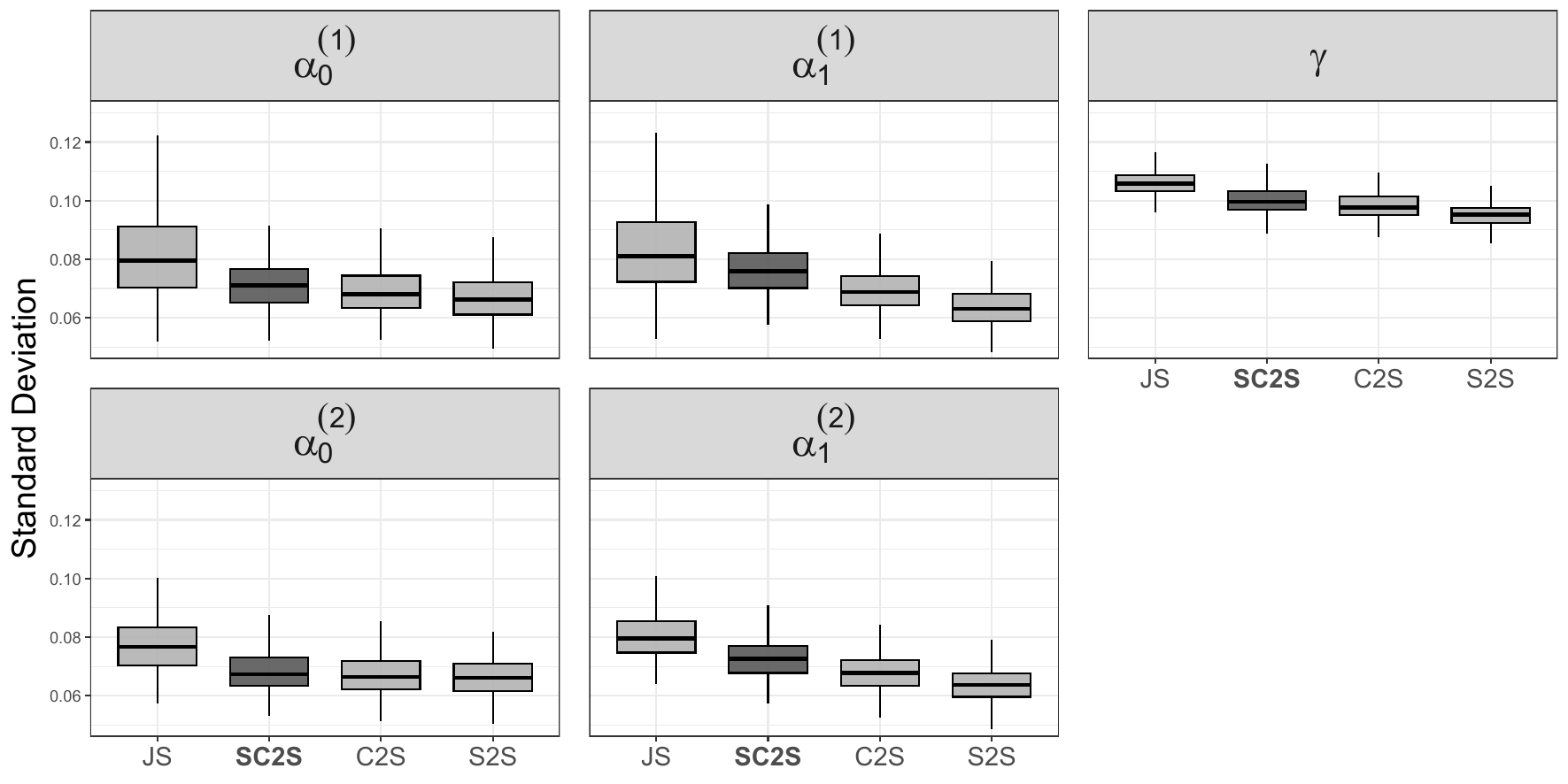}
    \caption{Posterior standard deviations for the survival parameters}
    \label{fig2}
\end{figure}

\noindent To examine the accuracy of the uncertainty estimation, Figure~\ref{fig2} examines the distribution of the posterior standard deviations across all simulations. As expected, the standard deviation from the S2S estimators is smaller than the competitors. This happens because this approach is structurally unable to propagate first-stage sampling error. However, the bias-corrected methods (C2S and SC2S) provide more realistic measures of uncertainty. It can be observed that the posterior dispersion obtained with our SC2S approach is closer to that of the fully joint specification (JS) than the S2S approach. This confirms that the SC2S strategy effectively propagates longitudinal uncertainty into the survival submodel, resulting in more reliable measures of precision. To conclude, across all survival-related parameters, SC2S approaches to the performance of the fully joint specification, while retaining a computationally efficient two-stage estimation structure. \\

\begin{figure}[H]
    \centering
    \includegraphics[width=0.7\linewidth]{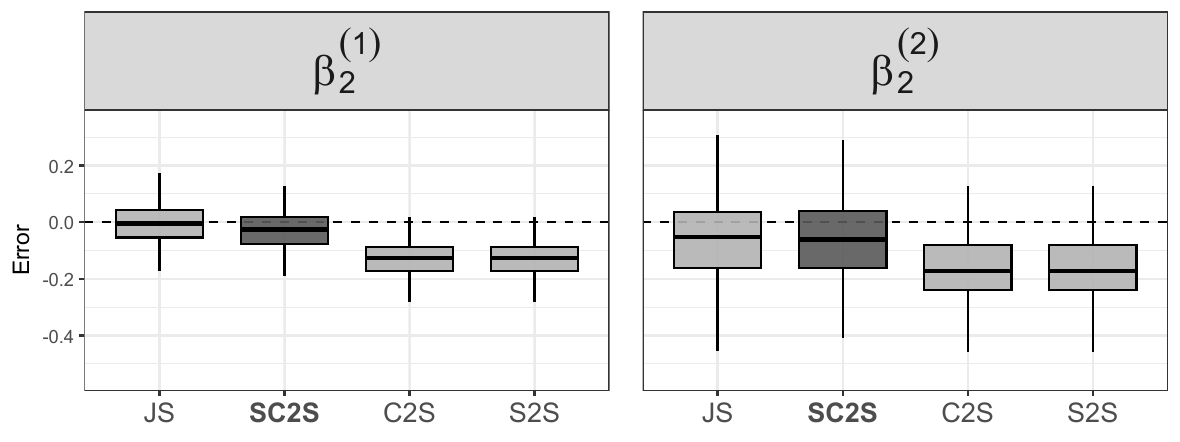}
    \caption{Estimation errors of the fixed time effect parameters}
    \label{fig3}
\end{figure}

\noindent Figure~\ref{fig3} focuses on the longitudinal parameters $\beta_2^{(1)}$ and $\beta_2^{(2)}$, and shows that they are affected by informative dropout under the S2S and C2S approaches. When the joint likelihood is not accounted for in the first stage, substantial bias arises in these slope fixed effects. Because neither method re-estimates them in the second stage, both S2S and C2S inherit this bias. In contrast, the SC2S approach explicitly re-estimates $(\beta_2^{(p)})_{p=1}^2$ in the second stage using the joint likelihood, leading to a substantial reduction in bias. The estimation of other longitudinal parameters (not shown here), however, does not exhibit notable first-stage bias for any of the methods.\\

\noindent To further assess the proximity of the proposed two-stage approaches to the full joint specification (JS), we computed Wasserstein distances between the posterior distributions of the model parameters obtained under each method. This distance provides a natural way to compare entire posterior distributions rather than point estimates, thereby capturing differences in both location and dispersion \citep{villani2009optimal}. For each model parameter, it can be interpreted as the minimal cost required to transform one posterior distribution into another. For each simulated dataset, the Wasserstein distance was computed for each parameter by comparing the posterior distributions obtained under JS and each competing approach, and then summarized across parameters. The results are reported in Table~\ref{tab2}, where distances are aggregated by parameter family, namely the longitudinal fixed effects $\{\beta_j^{(p)}\}$, the survival parameter $\gamma$, and the association parameters $\{\alpha_q^{(p)}\}$.

\begin{table}[H]
\centering 
\caption{Mean Wasserstein distances to JS by parameter family across 500 simulated datasets.}
\label{tab2}
\begin{tabular}{lccc}
\toprule
Family & JS--SC2S & JS--C2S & JS--S2S \\
\midrule
Beta             & 0.0555 & 0.0831 & 0.0831 \\
Gamma            & 0.0172 & 0.0198 & 0.0316 \\
Association      & 0.0409 & 0.0556 & 0.0830 \\
\bottomrule
\end{tabular}
\end{table}

\noindent Overall, as expected, the SC2S approach produces posterior distributions that are closest to those obtained under the full joint model, while the S2S approach exhibits the largest distances. The most substantial improvement is observed for the association parameters $\{\alpha_q^{(p)}\}$, with a reduction of approximately 51\% in Wasserstein distance when moving from S2S to SC2S. A more detailed parameter-by-parameter comparison is provided in Table~S2 (Supplementary Section S2). As anticipated, within the association family, the largest gains are observed for the slope-related parameters (i.e., $\alpha_1^{(p)}$), which are known to be particularly sensitive to informative dropout. Finally, threshold and discrimination parameters are not reported here, as their estimation is identical across the two-stage approaches.\\

\noindent The performance measures obtained with the SC2S approach are reported in Table~\ref{tab3} and in Table S1 (Supplementary Section~S2).  Across all classes of parameters, we observe very small estimation biases and coverage probabilities close to the nominal level using SC2S. In particular, the good performances obtained for the discrimination and threshold parameters confirm that the model successfully differentiates items and correctly identifies the ordering of response categories. For the longitudinal parameters, the fixed effects of both latent dimensions are accurately recovered. As discussed, the estimation accuracy for the coefficients $(\beta_2^{(p)})_{p=1}^2$ is slightly lower with SC2S than with the JS approach, but the bias is largely reduced compared with the classical two-stage approach. Finally, in the survival submodel, the association parameters are well recovered, demonstrating that the method effectively captures the distinct random effects associated with the two latent traits. \\

\begin{table}[!htbp]
\centering
\caption{Simulation results for $S = 500$ replications of sample of size $n=500$ with the SC2S approach : bias, root mean square error (RMSE) and effective coverage (COV) of 95\% credible intervals for the survival and latent processes parameters.}
\label{tab3}
\begin{tabular}{l c c c l}
\hline
Parameter & {True} & {Bias} & {RMSE} & {COV} \\
\hline
$\beta_0^{(1)}$ & \,\,0.50 & \,\,0.01 & 0.10 & 96\% \\
$\beta_0^{(2)}$ & \,\,1.00 & -0.03 & 0.14 & 98\% \\
$\beta_1^{(1)}$ & -1.00 & \,\,0.00 & 0.12 & 94\% \\
$\beta_1^{(2)}$ & -0.50 & \,\,0.06 & 0.20 & 99\% \\
$\beta_2^{(1)}$ & \,\,0.75 & -0.03 & 0.08 & 81\% \\
$\beta_2^{(2)}$ & \,\,0.75 & -0.06 & 0.16 & 75\% \\
\hline
$\gamma$ & \,\,0.50 & -0.01 & \,\,0.12 & 91\% \\
$\alpha_0^{(1)}$ & \,\,0.30 & \,\,0.03 & 0.10 & 88\% \\
$\alpha_1^{(1)}$ & \,\,0.30 & \,\,0.02 & 0.10 & 88\% \\
$\alpha_0^{(2)}$ & \,\,0.30 & \,\,0.01 & 0.07 & 96\% \\
$\alpha_1^{(2)}$ & \,\,0.30 & -0.01 & 0.07 & 93\% \\
\hline
\end{tabular}
\end{table}

\noindent The corresponding results under the current-value association structure are reported in Tables~S3 and~S4 (Supplementary Section~S3). The SC2S approach performs consistently in the current-value association setting, exhibiting very small estimation biases and coverage probabilities close to the nominal level. In this setting, the computation time is $12.9\; (8.6\text{--}26.9)$ minutes with the SC2S approach, compared with $ 30.4\; ( 21.3 - 49.5 )$ with the JS approach. \\

\noindent Taken together, these findings demonstrate that SC2S achieves a favourable balance between accuracy and computational efficiency, closely approximating the fully joint estimator while reducing runtime.

\section{Application to glioblastoma data}\label{sec5}

The dataset used in this study originates from a phase III clinical trial involving patients with first progression of glioblastoma. This trial was conducted by the European Organisation for Research and Treatment of Cancer (EORTC-26101 trial) to determine whether the combination of lomustine and bevacizumab (combination group) improved overall survival compared to lomustine alone (monotherapy group). Details of this trial can be found in \citet{clinic} (Identifier NCT01290939). While the primary endpoint of the trial is overall survival, these patients have a poor prognosis, making it crucial to evaluate the treatment's impact on their well-being.\\

\noindent To this end, quality-of-life data were collected longitudinally using the EORTC Quality of Life Questionnaires QLQ-C30 and QLQ-BN20. The latter module specifically assesses the effects of brain tumours and their treatment on symptoms, functioning, and health-related quality of life (HRQoL), whereas the former was developed for use in general cancer populations \citep{Fayers2001}.\\

\noindent Specifically, the QLQ-BN20 is composed of 20 ordinal items measuring different dimensions of quality of life including future uncertainty (four items), visual disorder (three items), motor dysfunction (three items), and communication deficit (three items). Seven additional single items assess headaches, seizures, drowsiness, hair loss, itchy skin, weakness of legs, and bladder control. Patients self-reported their responses at baseline and every 12 weeks thereafter until disease progression, allowing the monitoring of changes in HRQoL over time.\\

\noindent Trial results were previously published \citep{Wick} and did not demonstrate an overall survival advantage of lomustine plus bevacizumab over lomustine alone in patients with progressive glioblastoma, despite a prolonged progression-free survival (PFS). At that time, HRQoL was analysed using a sum-score approach and did not identify a significant treatment effect.\\

\noindent To demonstrate our methodology, we focus on the motor dysfunction (MD) and communication deficit (CD) scales, as these were identified as the primary domains of interest in Wick et al. (2017)\citep{Wick}. They are presented in Appendix A. Each item has a response range from 1 ("Not at all") to 4 ("Very much"), where higher responses indicate worse dysfunction or deficit. We impose identifiability constraints on one item from each dimension, that is, Item 1 and Item 4, by letting $d_{11} = d_{41} = 0$, diagonal elements of matrix $(\bm{a}_1,\bm{a}_4)^\top$ equal to 1 and off-diagonal elements equal to 0. \\

\noindent The dataset analysed includes $n = 423$ patients enrolled across multiple institutions in Europe. Of these, $145$ were randomized to receive monotherapy treatment, while the rest were randomized to receive a combination of lomustine and bevacizumab. Follow-up times for the longitudinal HRQoL data ranged from 1 to 96 weeks. The median overall survival was 40.4 weeks, and the median progression-free survival was 13.6 weeks. We consider progression-free survival as the event of interest for the survival component. Each patient completed the HRQoL questionnaire between 1 and 9 times during follow-up. \\

\noindent We consider the following joint model in our analysis:
\[
\left\lbrace
\begin{array}{l}
 h_{i}(t) = h_{0}(t) \exp \big\{ w_i\gamma + \sum_{p=1}^{2} ( \alpha^{(p)}_0 b_{i0}^{(p)} + \alpha^{(p)}_1 b_{i1}^{(p)} ) \big\},\\[0.1cm]
\eta^{(1)}_{i}(t) = \beta_0^{(1)} + \beta_1^{(1)} x_i + \beta_2^{(1)} t + b_{i0}^{(1)} + b_{i1}^{(1)} t, \\[0.1cm]
\eta^{(2)}_{i}(t) = \beta_0^{(2)} + \beta_1^{(2)} x_i + \beta_2^{(2)} t + b_{i0}^{(2)} + b_{i1}^{(2)} t, \\[0.1cm]
\mathrm{logit}\{\mathrm{P} (y_{ik}(t) \le l \mid \bm{\eta}_i(t)) \} = d_{kl} - (a^{(1)}_{k}\eta^{(1)}_{i}(t) + a^{(2)}_{k}\eta^{(2)}_{i}(t)),
\end{array}\right.
\]
where $\eta^{(1)}_{i}(t)$ and $\eta^{(2)}_{i}(t)$ represent the latent motor dysfunction and communication deficit of patient $i$ at time $t$. Here, $l = 1,\dots,4$ indicates response categories and $k = 1,\dots,6$ the six items. Higher latent scores correspond to worse functioning. We include a random intercept and a random slope for each latent dimension, resulting in $\bm{b_i} = (b_{i0}^{(1)}, b_{i1}^{(1)}, b_{i0}^{(2)}, b_{i1}^{(2)}) \sim N(0,\Sigma)$ with $\Sigma_{rr} = \sigma_r^2$ and $\Sigma_{rs} = \rho_{rs}\sigma_r\sigma_s$ for $r,s = 1,\dots,4$, $r < s$. In this application, the same treatment variable is included in both $\bm{w}_i$ and $\bm{x}_i$. \\

\noindent We use the slope-corrected two-stage (SC2S) approach to estimate this joint model. Estimation required approximately 9 minutes. Trace plots revealed no discernible trends in the MCMC chains, indicating adequate convergence. Furthermore, the Gelman–Rubin diagnostic \citep{gelamn} confirmed that the potential scale reduction factor ($R$) for all parameters remained below 1.1. 

\subsection{Results overview}
Estimated parameters for the latent and survival processes are reported in Table~\ref{tab4}. The results for the GRM parameters and for the covariance matrix are presented in Tables~S5 and~S6 in Supplementary Section~S4.\\

\begin{table}
 \sf\centering 
\caption{Estimation results for latent processes and survival submodel parameters : Posterior means and 95\% credible intervals (CI) are reported.}
\label{tab4}
\begin{tabular}{lcc}
\hline
 Parameter & Est. & CI 95\% \\
\hline
\multicolumn{3}{l}{\textbf{Motor dysfunction}}\\
$\beta_0^{(1)}$  \hspace{0.5cm}  & -1.530 & [-1.983;\:\,-1.137] \\
$\beta_1^{(1)}$    & \;\,0.606 & [\;\,0.187;\:\,1.073] \\
$\beta_2^{(1)}$  & \;\,0.010 & [-0.011;\:\,0.031] \\[0.2cm]

\multicolumn{3}{l}{\textbf{Communication deficit}}\\
$\beta_0^{(2)}$ & \;\,0.925 & [\;\,0.131;\:\,1.822] \\
$\beta_1^{(2)}$   & -0.425 & [-1.379;\:\,0.511] \\
$\beta_2^{(2)}$  & \;\,0.060 & [\;\,0.024;\:\,0.098] \\[0.2cm]

\multicolumn{3}{l}{\textbf{Progression-free survival}}\\
$\gamma$              & -1.411 & [-2.123;\:\,-0.882] \\
$\alpha_0^{(1)}$   & -0.286 & [-0.759;\:\,0.075] \\
$\alpha_1^{(1)}$  &-3.760 & [-24.79;\:\,21.09] \\
$\alpha_0^{(2)}$   & \;\,0.167 & [-0.027;\:\,0.396] \\
$\alpha_1^{(2)}$  & \,18.54 & [\;\,8.88;\:\,32.80] \\
\hline
\end{tabular}
\end{table}
\noindent First, on the longitudinal dimensions, Table \ref{tab4} reveals that treatment has a significant effect on the level of motor dysfunction. The positive coefficient ($\beta_1^{(1)} = 0.606$), with the monotherapy group as reference and given that higher latent scores reflect worse functioning, indicates that patients in the combination group have, on average, worse motor functioning. By contrast, treatment did not demonstrate a significant effect on communication deficit. Over time (measured in weeks), communication deficit deteriorated significantly ($\beta_2^{(2)} = 0.060$), whereas the time trend in motor dysfunction was non-significant ($\beta_2^{(1)} = 0.010$).\\

\noindent In the survival submodel, we observed a statistically significant treatment effect on progression-free survival ($\gamma=-1.411$). The association parameters linking individual deviations in the latent processes to the risk of progression highlight communication as the most prognostic dimension. In particular, the association with the random slope of communication deficit is large and significant ($\alpha_{1}^{(2)} = 18.54$). The magnitude of this estimate should be interpreted in light of the small between-patient variability of the communication slopes (Table S6: $\sigma_{4} = 0.079$). To improve the interpretability of this effect, we considered standardised association parameters. We obtained $\alpha_{1,\text{std}}^{(2)} = \alpha_{1}^{(2)} \times \sigma_{4} = 1.47$, which represents the log-hazard increase associated with a one standard-deviation increase in the communication random slope. In comparison, the standardised effect for the motor slope was much smaller and non-significant ($\alpha_{1,\text{std}}^{(1)} = \alpha_{1}^{(1)} \times \sigma_{2} = -0.14$). We therefore observed that patients exhibiting faster deterioration in communication abilities tend to have a higher risk of progression. Awareness of this relationship may encourage clinicians to monitor communication abilities carefully, as an accelerated decline may be indicative of disease progression. Such changes may prompt clinicians to arrange additional assessments (e.g., imaging) or adjust follow-up intensity. Furthermore, supportive interventions may be considered to help patients manage communication difficulties associated with disease progression (e.g., speech therapy).\\

\noindent At the measurement level, the graded response model behaved as expected and clearly recovered the two latent dimensions. It successfully identified the appropriate items loading on each component. Specifically, for all items, the highest and statistically significant discrimination parameters corresponded to their expected latent domain (i.e., motor dimension for Items 1–3 and communication dimension for Items 4–6).  In detail, discrimination parameters for items 1, 2, and 3 were concentrated on the motor factor (Component 1), with Item 3 (“unsteady on your feet”) showing the highest discrimination ($a^{(1)}_3=1.656$), while cross-loadings on the communication factor were negligible (e.g., $a^{(2)}_3\approx 0$). For communication items (Component 2), Item 5 (“difficulty speaking”) showed the strongest discrimination ($a^{(2)}_5=1.124$). Also, Item 2 shows a small but significant cross-loading, with a stronger contribution to the motor component.\\

\noindent Figure~\ref{plotapp} presents the estimated distribution of response probabilities for Items~3 and 5, identified as the most discriminative items for $\eta^{(1)}$ and $\eta^{(2)}$, respectively. It illustrates the temporal evolution of the probabilities that an average patient selects a given response category $l \in {1,\dots,4}$ at various time points. For Item~3 (motor domain), a clear treatment effect is visible, with consistently worse predicted response distributions under combination therapy. Item~5 (communication domain) shows a pronounced deterioration over time, consistent with the significant positive slope estimated for the communication latent process. Across items, higher values of the latent processes align with increasing probabilities of selecting more severe response categories, as expected under the graded response model. The cumulative hazard curves in the bottom panels show that higher progression risk is associated with higher communication decline (e.g., patients increasingly choosing the “Quite a bit’’ category for Item~5). \\

\noindent Finally, the random-effects covariance structure indicates meaningful associations between the two domains, both at baseline and in longitudinal change (Table S6). Random intercepts for motor and communication were positively correlated ($\rho_{13}=0.403$), suggesting that patients with worse baseline motor function tended to have worse baseline communication. Similarly, the random slopes were positively correlated ($\rho_{24}=0.527$), indicating that patients who deteriorated faster in motor function also tended to deteriorate faster in communication.

\section{Discussion}\label{sec6}

In this work, we proposed a slope‐corrected two‐stage (SC2S) approach for the joint analysis of multivariate ordinal longitudinal data and time‐to‐event outcomes within a multidimensional latent trait framework. This methodology provides a computationally efficient tool for the analysis of HRQoL data in oncology and other clinical research settings where multidimensional patient‐reported outcomes and terminal event are present. \\

\noindent Our method extends the corrected two-stage approach proposed by \citet{alvares2023}, which uses estimates from the longitudinal submodel to define informative prior distributions for the random effects in the survival submodel. In contrast to their work, we apply this strategy within a substantially more complex modelling framework, involving multivariate ordinal outcomes and multidimensional latent processes. We further extend the C2S by allowing longitudinal slope parameters to be re-estimated in the second stage, thereby correcting the bias introduced by informative dropout.\\

\noindent Through simulation studies, we demonstrated that both SC2S and C2S improve the estimation performance compared with the standard two‐stage approach. In particular, SC2S further reduces bias in survival association parameters and in longitudinal slope parameters. Importantly, its performance approaches that of fully joint Bayesian estimation (JS), while offering considerable reductions in computation time. We also evaluated the quality of uncertainty quantification through the posterior standard deviations. The results show that the standard two-stage approach tends to underestimate variability, as expected. In contrast, both C2S and SC2S provide more realistic measures of uncertainty. This highlights the importance of accounting for the variability of the latent trajectories when estimating survival parameters. To further assess the proximity to the fully joint model, we compared posterior distributions using the Wasserstein distance, which captures differences in both location and dispersion. The results show that SC2S produces posterior distributions closest to those obtained under JS.\\

\noindent Beyond these performance results, some practical considerations regarding model specification deserve attention. In this approach, accurate estimation of the random-effects covariance matrix \(\Sigma\) is crucial, as it directly determines the prior distribution of the random effects used in the second stage, and its misspecification may propagate through the model. Careful prior specification for \(\Sigma\) is therefore essential. Indeed, the impact of the prior on \(\Sigma\) depends on the amount of longitudinal information available. With few observations per subject, the prior can have a substantial influence, making robust prior specification especially important. Both specifications described in Section~\ref{priors} are valid. However, in practice, the LKJ-based parameterization is generally preferred, as it ensures a proper correlation structure and provides more stable inference, especially in higher dimensions.\\

\noindent The methodology was applied to longitudinal HRQoL data from a phase III glioblastoma trial. The proposed joint model identified clinically relevant treatment effects and associations between latent functional dimensions and progression‐free survival. In particular, the survival analysis highlighted communication decline as a prognostic marker: patients whose communication abilities deteriorate more rapidly are at increased risk of progression, as reflected by a significant association with the communication random slope. Moreover, the graded response model accurately recovered the expected latent structure of the HRQoL items. The model also captured meaningful correlation patterns between motor and communication deficits, indicating that patients starting worse or declining faster in one domain tended to do so in the other as well. These findings show the ability of the proposed approach to uncover clinically relevant patterns in multidimensional patient‐reported outcomes and to detect subtle changes that may not be detected by univariate or traditional score‐based analyses.\\

\noindent Finally, an advantage of the corrected two-stage approaches is that they can be easily generalised to many settings. They can accommodate a wide range of longitudinal models (including flexible trajectories and different outcome types) as well as more complex survival models such as competing-risks or multistate models. Elaborate association structures could also be considered.

\section*{Acknowledgments}
The authors acknowledge the support of the ARC project IMAL (grant 20/25-107) financed by the Wallonia-Brussels Federation and granted by the Académie universitaire Louvain. The authors thank the European Organization for Research and Treatment of Cancer for permission to use the data from EORTC-26101. The contents of this publication and methods used are solely the responsibility of the authors and do not necessarily represent the official views of the EORTC.

\subsection*{Conflict of interest}
The authors declare that they have no conflict of interest.
\subsection*{Data availability}
The data that support the findings of this study are not publicly available because they are owned by a third party (EORTC). Requests to access these data should be directed to the EORTC.

\subsection*{Ethics approval and consent to participate}
Ethical approval was obtained from the EORTC. Informed consent was obtained from all participants included in the study.

\subsection*{Funding}
This research was funded by the ARC project IMAL (grant 20/25-107).

\section*{Supporting information}

The following supporting information is available as part of the online article:

\noindent
\textbf{S1.}
{Distributional assumptions}
\\
\textbf{S2.}
{Simulation results for the random-effects (RE) association}
\\
\textbf{S3.}
{Simulation results for the current-value (CV) association}
\\
\textbf{S4.}
{Additional results from the application}

\section*{Appendix}

\appendix
\setcounter{figure}{0}
\setcounter{table}{0}
\renewcommand{\thefigure}{\thesection\arabic{figure}}
\renewcommand{\thetable}{\thesection\arabic{table}}

\section{BN20 scales\label{anx1}}
\textbf{{Motor dysfunction (MD)}}\\
\noindent Item 1: \textit{Did you have weakness on one side of your body?}\\
\noindent Item 2: \textit{Did you have trouble with your coordination?}\\
\noindent Item 3: \textit{Did you feel unsteady on your feet? }\\[0.2cm]
\textbf{{Communication deficit (CD)}}\\
\noindent Item 4: \textit{Did you have trouble finding the right words to express yourself?}\\
\noindent Item 5: \textit{Did you have difficulty speaking?}\\
\noindent Item 6: \textit{Did you have trouble communicating your thoughts?}\\[0.2cm]
\noindent Possible answers : 1 = "Not at All", \,2="A Little", \,3="Quite a Bit", \,4="Very Much"

\newpage
\section{Graphical results for the application}
\begin{figure}[!htbp]
    \centering
      \includegraphics[width=\textwidth]{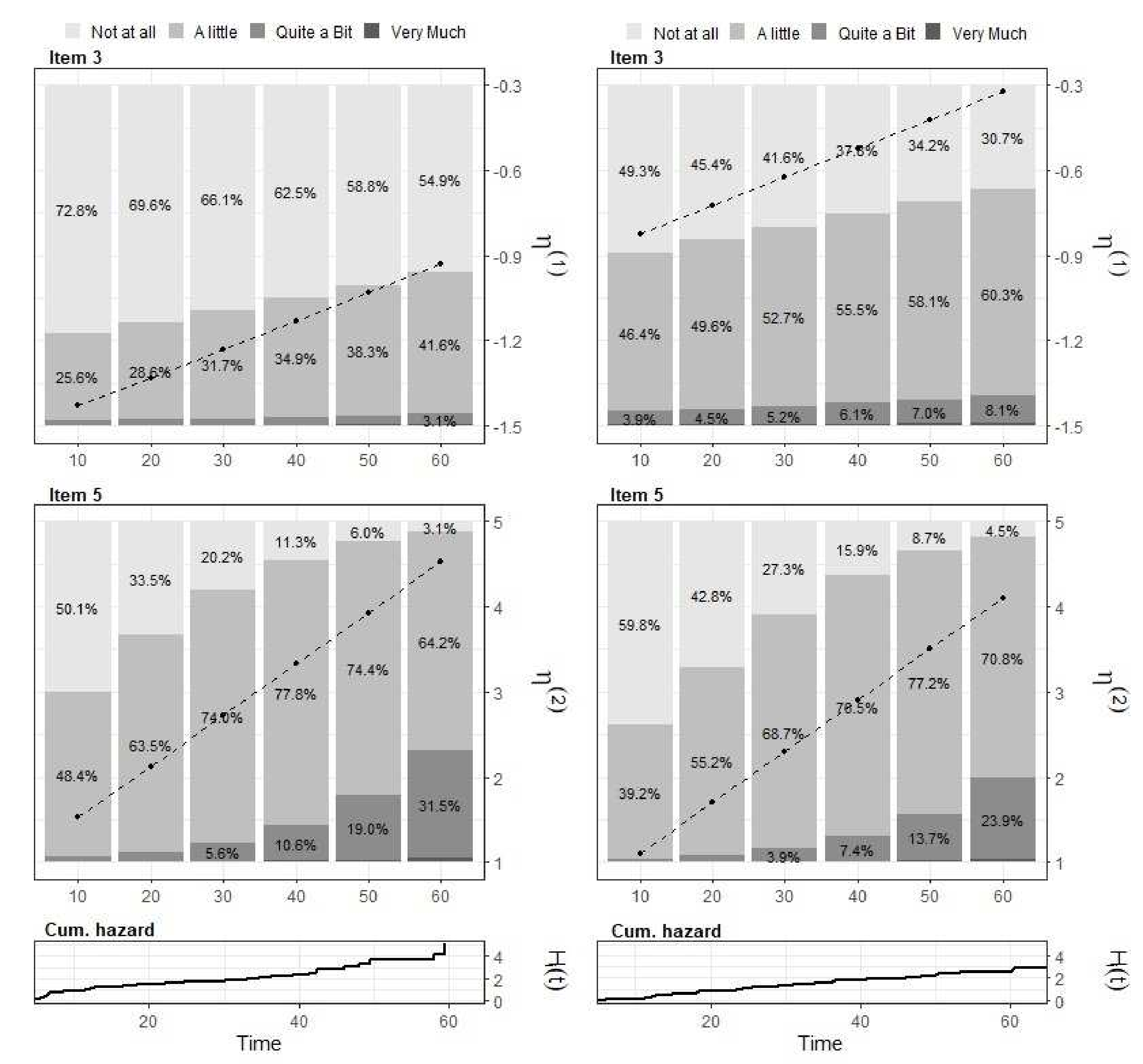}
\caption{
Predicted response probabilities over time for Item~3 (top) and Item~5 (middle) under monotherapy (left) and combination therapy (right). Stacked bars represent response probabilities, dashed lines the expected latent trajectories. Bottom panels show cumulative hazard functions.
}
    \label{plotapp}
\end{figure}

\newpage
\bibliography{HortenseDomsRef.bib}

\newpage\begin{center}
{\LARGE\bfseries Web-based Supporting Materials for ``A Bias-Corrected Two-Stage Approach for Joint Modelling of Multidimensional Longitudinal HRQoL and Survival Data''\par}
\end{center}

\vspace{0.5cm}
\renewcommand{\thesection}{S\arabic{section}}

\renewcommand{\thetable}{S\arabic{table}}
\setcounter{table}{0}
\renewcommand{\thefigure}{S\arabic{figure}}
\setcounter{figure}{0}
\setcounter{section}{0}
\section{Distributional assumptions}\label{3.C1}
\begin{equation*}
\label{eq3}
\begin{split}
p(T_i, \delta_{i} \mid \bm{b}_i ; \theta_s)
&= \big[ h_{i}(T_i \mid \bm{b}_i ; \theta_s) \big]^{\delta_{i}} \,
   S_{i}(T_i \mid \bm{b}_i ; \theta_s) \\[0.15cm]
&= \Big[ h_{0}(T_i)
   \exp \{ \bm{\gamma}^\top \bm{w}_i + \alpha\, m(T_i, \bm{b}_i) \} \Big]^{\delta_{i}} \\[0.15cm]
&\quad \times
\exp \Bigg(
 - \int_0^{T_i} h_{0}(s)
 \exp \{ \bm{\gamma}^\top \bm{w}_i + \alpha\, m(s, \bm{b}_i) \} \, ds
\Bigg) \\[0.15cm]
&= \Bigg[
\exp \Big\{
\sum_{u=1}^{U} \gamma_{h_0,u} B_u(T_i)
+ \bm{\gamma}^\top \bm{w}_i
+ \alpha\, m(T_i, \bm{b}_i)
\Big\}
\Bigg]^{\delta_{i}} \\[0.15cm]
&\quad \times
\exp \Bigg[
- \exp\!\big\{ \bm{\gamma}^\top \bm{w}_i \big\}
\int_0^{T_i}
\exp \Big\{
\sum_{u=1}^{U} \gamma_{h_0,u} B_u(s)
+ \alpha\, m(s, \bm{b}_i)
\Big\} ds
\Bigg]
\end{split}
\end{equation*}

\begin{equation*}
\label{eq:grm_pmf_new}
\begin{split}
p\bigl(y_{ik}(t) \mid \bm{b}_i ; \theta_y \bigr)
&= \prod_{l=1}^{L_k}
\Pr\!\bigl(y_{ik}(t)=l \mid \bm{\eta}_i(t)\bigr)^{\mathbbm{1}_{\{y_{ik}(t)=l\}}} \\[0.2cm]
&= \Big\{ \mathrm{expit}\!\big(d_{k1}-\bm{a}_k^{\top}\bm{\eta}_i(t)\big)
\Big\}^{\mathbbm{1}_{\{y_{ik}(t)=1\}}} \\[0.15cm]
&\quad \times \prod_{l=2}^{L_k-1}
\Big\{
\mathrm{expit}\!\big(d_{kl}-\bm{a}_k^{\top}\bm{\eta}_i(t)\big)
-
\mathrm{expit}\!\big(d_{k,l-1}-\bm{a}_k^{\top}\bm{\eta}_i(t)\big)
\Big\}^{\mathbbm{1}_{\{y_{ik}(t)=l\}}} \\[0.15cm]
&\quad \times
\Big\{
1-\mathrm{expit}\!\big(d_{k,L_k-1}-\bm{a}_k^{\top}\bm{\eta}_i(t)\big)
\Big\}^{\mathbbm{1}_{\{y_{ik}(t)=L_k\}}}
\end{split}
\end{equation*}

\noindent where $\text{expit}(x) = \frac{1}{1 + e^{-x}}$. \\[0.2cm]

\begin{equation*}
\label{eq5}
p(\bm{b}_i ; \theta_b)
= (2\pi)^{-q/2} \det({\Sigma})^{-1/2}
\exp\!\left(
-\tfrac{1}{2}\bm{b}_i^{\top}{\Sigma}^{-1}\bm{b}_i
\right)
\end{equation*}

\noindent  where $q$ denotes the dimension of the random effect vector. \vspace{0.3cm}

\newpage
\section{Simulation results for the random-effects (RE) association}\label{3.C2}
\begin{table}[H]
\centering
\caption{Bias, RMSE, and 95\% coverage for discrimination (top) and threshold (bottom) parameters under the SC2S approach with RE association. Parameters marked “/” were fixed for identifiability and therefore not estimated.}\label{sim-a}
\begin{subtable}{0.48\linewidth}
\centering
\begin{tabular}{l c c c l}
\toprule
 & {True} & {Bias} & {RMSE} & {COV} \\
\midrule
$a_1^{(1)}$  & 1.00 & / & / & / \\
$a_2^{(1)}$  & 0.10 & 0.07 & 0.18 & 98\% \\
$a_3^{(1)}$  & 1.50 & 0.05 & 0.13 & 95\% \\
$a_4^{(1)}$  & 2.00 & 0.05 & 0.13 & 95\% \\
$a_5^{(1)}$  & 0.10 & 0.11 & 0.29 & 99\% \\
$a_6^{(1)}$  & 1.20 & 0.04 & 0.09 & 95\% \\
$a_7^{(1)}$  & 1.60 & 0.05 & 0.13 & 96\% \\
$a_8^{(1)}$  & 0.20 & 0.08 & 0.22 & 99\% \\
$a_9^{(1)}$  & 0.30 & 0.07 & 0.19 & 98\% \\
$a_{10}^{(1)}$& 0.40 & 0.11 & 0.28 & 98\% \\
\bottomrule
\end{tabular}
\end{subtable}
\hfill
\begin{subtable}{0.48\linewidth}
\centering
\begin{tabular}{l c c c l}
\toprule
 & {True} & {Bias} & {RMSE} & {COV} \\
\midrule
$a_1^{(2)}$  & 0.00 & / & / & / \\
$a_2^{(2)}$  & 1.00 & / & / & / \\
$a_3^{(2)}$  & 0.40 & \;\,0.01 & 0.05 & 95\% \\
$a_4^{(2)}$  & 0.10 & -0.00 & 0.06 & 95\% \\
$a_5^{(2)}$  & 1.60 & \;\,0.02 & 0.10 & 95\% \\
$a_6^{(2)}$  & 0.20 & \;\,0.00 & 0.04 & 95\% \\
$a_7^{(2)}$  & 0.30 & \;\,0.00 & 0.05 & 95\% \\
$a_8^{(2)}$  & 1.20 & \;\,0.01 & 0.07 & 92\% \\
$a_9^{(2)}$  & 1.00 & \;\,0.01 & 0.06 & 95\% \\
$a_{10}^{(2)}$& 1.50 & \;\,0.03 & 0.09 & 92\% \\
\bottomrule
\end{tabular}
\end{subtable}
\\[0.5cm]
\begin{subtable}{0.48\linewidth}
\centering
\begin{tabular}{l c c c l}
\toprule
 & {True} & {Bias} & {RMSE} & {COV} \\
\midrule
$d_{1,1}$  & 0.00 & / & / & / \\
$d_{2,1}$  & 0.00 & / & / & / \\
$d_{3,1}$  & 1.00 & 0.03 & 0.14 & 96\% \\
$d_{4,1}$  & 2.00 & 0.04 & 0.19 & 95\% \\
$d_{5,1}$  & 0.80 & 0.01 & 0.15 & 97\% \\
$d_{6,1}$  & 0.50 & 0.02 & 0.13 & 90\% \\
$d_{7,1}$  & 0.30 & 0.02 & 0.14 & 97\% \\
$d_{8,1}$  & 0.60 & 0.01 & 0.12 & 96\% \\
$d_{9,1}$  & 1.20 & 0.02 & 0.11 & 95\% \\
$d_{10,1}$ & 1.50 & 0.03 & 0.16 & 94\% \\
\midrule
$d_{1,3}$  & 2.50 & -0.00 & 0.10 & 95\% \\
$d_{2,3}$  & 2.70 & 0.01 & 0.10 & 95\% \\
$d_{3,3}$  & 3.50 & 0.05 & 0.19 & 94\% \\
$d_{4,3}$  & 4.50 & 0.08 & 0.25 & 95\% \\
$d_{5,3}$  & 3.30 & 0.03 & 0.19 & 95\% \\
$d_{6,3}$  & 3.00 & 0.06 & 0.17 & 91\% \\
$d_{7,3}$  & 2.80 & 0.05 & 0.16 & 94\% \\
$d_{8,3}$  & 3.10 & 0.02 & 0.15 & 96\% \\
$d_{9,3}$  & 3.70 & 0.03 & 0.15 & 96\% \\
$d_{10,3}$ & 4.00 & 0.06 & 0.22 & 93\% \\
\bottomrule
\end{tabular}
\end{subtable}
\hfill
\begin{subtable}{0.48\linewidth}
\centering
\begin{tabular}{l c c c l}
\toprule
 & {True} & {Bias} & {RMSE} & {COV} \\
\midrule
$d_{1,2}$  & 1.00 & 0.00 & 0.07 & 95\% \\
$d_{2,2}$  & 1.00 & 0.01 & 0.06 & 96\% \\
$d_{3,2}$  & 2.00 & 0.04 & 0.16 & 95\% \\
$d_{4,2}$  & 3.00 & 0.05 & 0.21 & 97\% \\
$d_{5,2}$  & 1.80 & 0.02 & 0.16 & 96\% \\
$d_{6,2}$  & 1.50 & 0.03 & 0.14 & 94\% \\
$d_{7,2}$  & 1.30 & 0.02 & 0.14 & 97\% \\
$d_{8,2}$  & 1.60 & 0.01 & 0.13 & 94\% \\
$d_{9,2}$  & 2.20 & 0.02 & 0.12 & 97\% \\
$d_{10,2}$ & 2.50 & 0.04 & 0.18 & 94\% \\
\midrule
$d_{1,4}$  & 5.00 & 0.02 & 0.18 & 95\% \\
$d_{2,4}$  & 5.50 & 0.01 & 0.19 & 95\% \\
$d_{3,4}$  & 6.00 & 0.07 & 0.25 & 96\% \\
$d_{4,4}$  & 7.00 & 0.13 & 0.34 & 95\% \\
$d_{5,4}$  & 5.80 & 0.06 & 0.27 & 94\% \\
$d_{6,4}$  & 5.50 & 0.08 & 0.24 & 95\% \\
$d_{7,4}$  & 5.30 & 0.07 & 0.23 & 95\% \\
$d_{8,4}$  & 5.60 & 0.05 & 0.23 & 93\% \\
$d_{9,4}$  & 6.20 & 0.04 & 0.23 & 96\% \\
$d_{10,4}$ & 6.50 & 0.09 & 0.29 & 94\% \\
\bottomrule
\end{tabular}
\end{subtable}
\end{table}

\newpage

\begin{table}[ht]
\centering
\caption{Mean Wasserstein distances to the joint specification (JS) across 500 simulated datasets for selected survival and latent process parameters.}
\label{distance_full}
\begin{tabular}{lccc}
\toprule
Parameter & JS--SC2S & JS--C2S & JS--S2S \\
\midrule
$\beta_0^{(1)}$ & 0.0085 & 0.0085 & 0.0085 \\
$\beta_0^{(2)}$ & 0.0681 & 0.0681 & 0.0681 \\
$\beta_1^{(1)}$ & 0.0114 & 0.0114 & 0.0114 \\
$\beta_1^{(2)}$ & 0.1396 & 0.1396 & 0.1396 \\
$\beta_2^{(1)}$ & 0.0359 & 0.1225 & 0.1225 \\
$\beta_2^{(2)}$ & 0.0692 & 0.1487 & 0.1487 \\
\midrule
$\gamma$ & 0.0172 & 0.0198 & 0.0316 \\
$\alpha_0^{(1)}$ & 0.0648 & 0.0673 & 0.0689 \\
$\alpha_1^{(1)}$ & 0.0642 & 0.0836 & 0.1338 \\
$\alpha_0^{(2)}$ & 0.0172 & 0.0209 & 0.0203 \\
$\alpha_1^{(2)}$ & 0.0174 & 0.0507 & 0.1092 \\
\bottomrule
\end{tabular}
\end{table}

\section{Simulation results for the current-value (CV) association}\label{3.C3}
\begin{table}[H]
\centering
\caption{Bias, RMSE, and 95\% coverage for the survival and latent processes parameters under the SC2S approach with CV association.}
\label{sim-surv-curr}
\begin{tabular}{l c c c l}
\toprule
Parameter & {True} & {Bias} & {RMSE} & {COV} \\
\midrule
$\beta_0^{(1)}$ & -0.50 & -0.00 & 0.08 & 94\% \\
$\beta_0^{(2)}$ & -0.30 & \;\,0.00 & 0.09 & 92\% \\
$\beta_1^{(1)}$ & \;\,0.10 & -0.00 & 0.08 & 92\% \\
$\beta_1^{(2)}$ & \;\,0.15 & \;\,0.00 & 0.08 & 92\% \\
$\beta_2^{(1)}$ & \;\,0.50 & -0.01 & 0.09 & 94\% \\
$\beta_2^{(2)}$ & \;\,0.75 & -0.02 & 0.10 & 90\% \\
\midrule
$\gamma$ & \;\,0.30 & \;\,0.01 & 0.09 & 92\% \\
$\alpha^{(1)}$ & \;\,0.15 & -0.05 & 0.08 & 89\% \\
$\alpha^{(2)}$ & \;\,0.10 & -0.03 & 0.07 & 92\% \\
\bottomrule
\end{tabular}
\end{table}

\newpage
\begin{table}[H]
\centering
\caption{Bias, RMSE, and 95\% coverage for discrimination (top) and threshold (bottom) parameters under the SC2S approach with CV association.}\label{sim-surv-curr-2}
\begin{subtable}{0.48\linewidth}
\centering
\begin{tabular}{l c c c l}
\toprule
 & {True} & {Bias} & {RMSE} & {COV} \\
\midrule
$a_1^{(1)}$  & 1.00 & / & / & / \\
$a_2^{(1)}$  & 0.00 & / & / & / \\
$a_3^{(1)}$  & 1.50 & \;\,0.08 & 0.19 & 95\% \\
$a_4^{(1)}$  & 2.00 & \;\,0.11 & 0.26 & 89\% \\
$a_5^{(1)}$  & 0.10 & \;\,0.02 & 0.15 & 95\% \\
$a_6^{(1)}$  & 1.20 & \;\,0.06 & 0.16 & 89\% \\
$a_7^{(1)}$  & 1.60 & \;\,0.06 & 0.20 & 94\% \\
$a_8^{(1)}$  & 0.20 & -0.00 & 0.13 & 92\% \\
$a_9^{(1)}$  & 0.30 & \;\,0.02 & 0.11 & 96\% \\
$a_{10}^{(1)}$& 0.40 & \;\,0.02 & 0.14 & 97\% \\
\bottomrule
\end{tabular}
\end{subtable}
\hfill
\begin{subtable}{0.48\linewidth}
\centering
\begin{tabular}{l c c c l}
\toprule
 & {True} & {Bias} & {RMSE} & {COV} \\
\midrule
$a_1^{(2)}$  & 0.00 & / & / & / \\
$a_2^{(2)}$  & 1.00 & / & / & / \\
$a_3^{(2)}$  & 0.40 & -0.03 & 0.11 & 94\% \\
$a_4^{(2)}$  & 0.10 & -0.05 & 0.16 & 92\% \\
$a_5^{(2)}$  & 1.60 & \;\,0.01 & 0.14 & 94\% \\
$a_6^{(2)}$  & 0.20 & -0.02 & 0.10 & 95\% \\
$a_7^{(2)}$  & 0.30 & -0.03 & 0.11 & 95\% \\
$a_8^{(2)}$  & 1.20 & \;\,0.02 & 0.12 & 92\% \\
$a_9^{(2)}$  & 1.00 & \;\,0.00 & 0.10 & 95\% \\
$a_{10}^{(2)}$& 1.50 & \;\,0.02 & 0.14 & 92\% \\
\bottomrule
\end{tabular}
\end{subtable}
\\[0.5cm]
\begin{subtable}{0.48\linewidth}
\centering
\begin{tabular}{l c c c l}
\toprule
 & {True} & {Bias} & {RMSE} & {COV} \\
\midrule
$d_{1,1}$  & 0.00 & / & / & / \\
$d_{2,1}$  & 0.00 & / & / & / \\
$d_{3,1}$  & 1.00 & -0.01 & 0.13 & 96\% \\
$d_{4,1}$  & 2.00 & -0.01 & 0.19 & 90\% \\
$d_{5,1}$  & 0.80 & \;\,0.01 & 0.14 & 95\% \\
$d_{6,1}$  & 0.50 & -0.01 & 0.11 & 94\% \\
$d_{7,1}$  & 0.30 & \;\,0.01 & 0.13 & 94\% \\
$d_{8,1}$  & 0.60 & \;\,0.02 & 0.11 & 95\% \\
$d_{9,1}$  & 1.20 & \;\,0.01 & 0.09 & 96\% \\
$d_{10,1}$ & 1.50 & \;\,0.03 & 0.15 & 94\% \\
\midrule
$d_{1,3}$  & 2.50 & \;\,0.00 & 0.08 & 97\% \\
$d_{2,3}$  & 2.70 & -0.00 & 0.09 & 94\% \\
$d_{3,3}$  & 3.50 & -0.01 & 0.17 & 94\% \\
$d_{4,3}$  & 4.50 & -0.00 & 0.23 & 95\% \\
$d_{5,3}$  & 3.30 & \;\,0.02 & 0.18 & 95\% \\
$d_{6,3}$  & 3.00 & \;\,0.01 & 0.12 & 97\% \\
$d_{7,3}$  & 2.80 & -0.00 & 0.16 & 94\% \\
$d_{8,3}$  & 3.10 & \;\,0.03 & 0.14 & 96\% \\
$d_{9,3}$  & 3.70 & \;\,0.01 & 0.13 & 95\% \\
$d_{10,3}$ & 4.00 & \;\,0.04 & 0.19 & 91\% \\
\bottomrule
\end{tabular}
\end{subtable}
\hfill
\begin{subtable}{0.48\linewidth}
\centering
\begin{tabular}{l c c c l}
\toprule
 & {True} & {Bias} & {RMSE} & {COV} \\
\midrule
$d_{1,2}$  & 1.00 & \;\,0.00 & 0.05 & 97\% \\
$d_{2,2}$  & 1.00 & \;\,0.00 & 0.05 & 92\% \\
$d_{3,2}$  & 2.00 & -0.00 & 0.14 & 98\% \\
$d_{4,2}$  & 3.00 & -0.00 & 0.21 & 94\% \\
$d_{5,2}$  & 1.80 & \;\,0.01 & 0.15 & 92\% \\
$d_{6,2}$  & 1.50 & \;\,0.00 & 0.10 & 97\% \\
$d_{7,2}$  & 1.30 & -0.01 & 0.14 & 92\% \\
$d_{8,2}$  & 1.60 & \;\,0.02 & 0.12 & 95\% \\
$d_{9,2}$  & 2.20 & \;\,0.01 & 0.11 & 97\% \\
$d_{10,2}$ & 2.50 & \;\,0.03 & 0.16 & 95\% \\
\midrule
$d_{1,4}$  & 5.00 & -0.02 & 0.18 & 98\% \\
$d_{2,4}$  & 5.50 & \;\,0.04 & 0.20 & 92\% \\
$d_{3,4}$  & 6.00 & \;\,0.01 & 0.26 & 92\% \\
$d_{4,4}$  & 7.00 & \;\,0.03 & 0.35 & 94\% \\
$d_{5,4}$  & 5.80 & \;\,0.04 & 0.24 & 93\% \\
$d_{6,4}$  & 5.50 & \;\,0.04 & 0.22 & 94\% \\
$d_{7,4}$  & 5.30 & -0.00 & 0.24 & 88\% \\
$d_{8,4}$  & 5.60 & \;\,0.06 & 0.21 & 95\% \\
$d_{9,4}$  & 6.20 & \;\,0.02 & 0.23 & 95\% \\
$d_{10,4}$ & 6.50 & \;\,0.05 & 0.26 & 91\% \\
\bottomrule
\end{tabular}
\end{subtable}
\end{table}

\newpage

\section{Additional results from the application}\label{3.C6}
\begin{table}[ht]
\small \sf\centering
\caption{Estimation results of GRM submodel parameters: posterior means and 95\% credible intervals (CI).}
\label{app_GRM}

\begin{minipage}[t]{0.48\linewidth}
\centering
\begin{tabular}{llcc}
\hline
 & & Est. & CI 95\% \\
\hline
$\mathrm{Item 1}$ & $d_{11}$ & \:\,0.000 & / \\
& $d_{12}$ & \:\,1.547 & [\:\,1.304;\:\,1.815] \\
& $d_{13}$ & \:\,2.833 & [\:\,2.455;\:\,3.247] \\
$\mathrm{Item 2}$ & $d_{21}$ & -0.416 & [-0.717;-0.153] \\
& $d_{22}$ & \:\,1.993 & [\:\,1.675;\:\,2.313] \\
& $d_{23}$ & \:\,3.949 & [\:\,3.448;\:\,4.514] \\
$\mathrm{Item 3}$ & $d_{31}$ & -1.406 & [-2.122;-0.866] \\
& $d_{32}$ & \:\,1.730 & [\:\,1.285;\:\,2.173] \\
& $d_{33}$ & \:\,4.185 & [\:\,3.523;\:\,5.042] \\
$\mathrm{Item 4}$ & $d_{41}$ & \:\,0.000 & / \\
& $d_{42}$ & \:\,4.244 & [\:\,3.633;\:\,4.964] \\
& $d_{43}$ & \:\,7.259 & [\:\,6.268;\:\,8.400] \\
$\mathrm{Item 5}$ & $d_{51}$ & \:\,1.518 & [\:\,0.988;\:\,2.101] \\
& $d_{52}$ & \:\,5.678 & [\:\,4.732;\:\,6.898] \\
& $d_{53}$ & \:\,9.376 & [\:\,7.917;\:\,11.25] \\
$\mathrm{Item 6}$ & $d_{61}$ & \:\,0.664 & [\:\,0.264;\:\,1.081] \\
& $d_{62}$ & \:\,4.559 & [\:\,3.887;\:\,5.332] \\
& $d_{63}$ & \:\,7.618 & [\:\,6.562;\:\,8.816] \\
\hline
\end{tabular}
\end{minipage}
\hfill
\begin{minipage}[t]{0.48\linewidth}
\centering
\begin{tabular}{lcc}
\hline
 & Est. & CI 95\% \\
\hline
$\mathrm{Dimension\,1}$ &  & \\[0.15cm]
$a_1^{(1)}$ & \:\,1.000 & / \\
$a_2^{(1)}$ & \:\,0.787 & [\:\,0.583;1.007] \\
$a_3^{(1)}$ & \:\,1.656 & [\:\,1.134;2.383] \\
$a_4^{(1)}$ & \:\,0.000 & / \\
$a_5^{(1)}$ & \:\,0.140 & [-0.057;0.330] \\
$a_6^{(1)}$ & \:\,0.145 & [-0.025;0.321] \\[0.2cm]
$\mathrm{Dimension\,2}$ &  & \\[0.15cm]
$a_1^{(2)}$ & \:\,0.000 & / \\
$a_2^{(2)}$ & \:\,0.124 & [\:\,0.064;0.185] \\
$a_3^{(2)}$ & -0.014 & [-0.121;0.070] \\
$a_4^{(2)}$ & \:\,1.000 & / \\
$a_5^{(2)}$ & \:\,1.124 & [\:\,0.849;1.481] \\
$a_6^{(2)}$ & \:\,0.882 & [\:\,0.691;1.104] \\
\hline
\end{tabular}
\end{minipage}
\end{table}

\begin{table}[H]
\small \sf\centering 
\caption{Estimation results of covariance matrix parameters: estimates and 95\% credible intervals (CI) are presented.}
\label{app_cov}
\begin{tabular}{lcc}
\hline
 & Est. & CI 95\% \\
\hline
$\sigma_1$   & \:\,1.756 & [\:\,1.425;\:\,2.154] \\
$\sigma_2$   & \:\,0.038 & [\:\,0.014;\:\,0.065] \\
$\sigma_3$   & \:\,3.959 & [\:\,3.299;\:\,4.697] \\
$\sigma_4$   & \:\,0.079 & [\:\,0.049;\:\,0.112] \\
$\rho_{12}$  & \:\,0.058 & [-0.302;\:\,0.462] \\
$\rho_{13}$  & \:\,0.403 & [\:\,0.240;\:\,0.552] \\
$\rho_{14}$  & \:\,0.229 & [-0.079;\:\,0.521] \\
$\rho_{23}$  & -0.213 & [-0.553;\:\,0.106] \\
$\rho_{24}$  & \:\,0.527 & [\:\,0.100;\:\,0.832] \\
$\rho_{34}$  & -0.202 & [-0.513;\:\,0.185] \\
\hline
\end{tabular}
\end{table}

\end{document}